\documentclass{article}[12 pt]
\usepackage[dvips]{graphicx}
\title{Elliptic Function Representation of Doubly Periodic Two-Dimensional Stokes Flows}
\author{Mark A. Peterson, Danti Chen and Mengqi Ding\\
Mount Holyoke College\\ South Hadley, MA 01075 USA}

\usepackage[dvips]{graphicx}

\begin{document}
\maketitle
\begin{abstract}
We construct doubly periodic Stokes flows in two dimensions using elliptic functions.
This method has advantages when the doubly periodic lattice of obstacles has less than
maximal symmetry.  We find the mean flow through an arbitrary lattice
in response to a pressure gradient in an arbitrary direction,
and show in a typical example that the shorter of the
two period lattice vectors is an ``easy direction"
for the flow, an eigenvector of the conductance tensor corresponding to maximal conductance.
\end{abstract}

It is known, and we rederive it below, that two-dimensional (2D) Stokes
flows can be represented in terms of two complex analytic functions
\cite{HB}.  It is plausible, then, that doubly periodic 2D Stokes flows
should have a representation in terms of doubly periodic complex analytic
functions, that is, elliptic functions \cite{Lawden,AandS}. Such a
representation was promised in 1959 by H. Hasimoto \cite{Hasimoto1}, but
it did not appear. Other authors in the succeeding decades alluded to such
a representation \cite{Zaks}, and even, like Hasimoto, quoted results
following from it \cite{Acrivos1}.  The Hasimoto article may have appeared
much later as lecture notes in Japanese \cite{Hasimoto2}. Meanwhile, there
are other, perhaps more straightforward, ways to represent doubly periodic
2D Stokes flows.  These include matching of flows around a single obstacle
across periodic cell boundaries
\cite{Acrivos1,Bruschke,WangCY,Hellou,Kirsh}, integral equation methods
\cite{Greengard}, biharmonic solvers on a grid \cite{Zaks}, and finite
element methods.

In spite of this long history, we have thought it useful to present the
elliptic function method, because there is still, apparently, no readily
available description of it. Furthermore, as we shall show, this approach
solves one aspect of the problem which is not at all simple in the most
common cell matching approach, namely the appropriate boundary condition
for flow through a general periodic lattice in a general direction.  With
this method we describe the typical flow through a generic lattice.

\section{2D Stokes Flows}
Let us represent the 2D flow with velocity vector field $\vec{u}(x,y)$ as a complex
scalar function $u$ by means of the usual isomorphism
\begin{equation}
\vec{u}=u_x{\rm\bf \hat{x}}+u_y{\rm\bf \hat{y}}\quad\leftrightarrow\quad u(x,y)=u_x+iu_y
\end{equation}
Regarding the $x$-$y$ plane as the complex $z$ plane, we note that the divergence and curl
of $u$ are given by
\begin{eqnarray}
\label{divu}
{\rm div}\,\vec{u}&=&2\Re (\partial u/\partial z)=0\\
\label{curlu}
{\rm curl}\,\vec{u}&=&2\Im (\partial u/\partial z)=\omega
\end{eqnarray}
Here the first equation expresses incompressibility of the flow $u$,
and the second defines the vorticity $\omega$,
understood as the (scalar) component of a vector field normal to the plane.
Similarly, the gradient of the pressure $P$ in this complex representation is
\begin{equation}
\vec{\nabla}P\quad\leftrightarrow\quad 2\partial P/\partial \overline{z}
\end{equation}
The Stokes equation, representing the balance of viscous stress in the fluid by the pressure
gradient, then becomes
\begin{equation}
\mu\nabla^2 \vec{u}=\vec{\nabla}P\quad\leftrightarrow\quad 4\mu\,\partial^2 u/\partial \overline{z} \partial z\,
=2\partial P/\partial \overline{z}
\end{equation}
where $\mu$ is the viscosity.  In light of Eqs.~(\ref{divu}) and (\ref{curlu}) this means that
\begin{equation}
\partial(P/\mu-i\omega)/\partial\overline{z}=0
\end{equation}
that is, that
\begin{equation}
\label{functionf}
f(z)=P/\mu-i\omega
\end{equation}
is a single-valued function of $z$, holomorphic except for
possible poles \cite{HB}.  If we choose such a function $f$, we can integrate
Eqs.~(\ref{divu}) and (\ref{curlu}) to find the flow
\begin{equation}
\label{uformula}
u=\frac{1}{4}(z\overline{f}-\int^z f\,dz+\overline{g})\,
\end{equation}
where $g(z)$ is a second holomorphic function, having logarithmic
singularities at the poles of $f$.  Thus $u$ is represented
in terms of two holomorphic functions, $f$ and $g$.

The force exerted by the flow $u$ on a finite
obstacle, given by closed contour $C$,
just involves the enclosed residues of $f$.  The force on a small, oriented
line segment $\Delta z$ due to the fluid on its right is
\begin{equation}
\Delta F=iP\Delta z+2i\mu (\partial u/\partial \overline{z})\Delta\overline{z}
\end{equation}
Now, using Eqs.~(\ref{functionf}) and (\ref{uformula}), and
integrating over the closed curve $C$, oriented in
the conventional positive direction, we find the force on $C$ due to the fluid outside it
\begin{eqnarray}
F&=&\frac{i\mu}{2}\left[\oint_C(f+\overline{f})\,dz+\oint_C(z\overline{f'}+\overline{g'})\,d\overline{z}\right]\\
&=&\frac{i\mu}{2}\left[\oint_Cd(z\overline{f}+\overline{g})+\oint_C f\,dz\right]\\
\label{FonObstacle}
&=&i\mu\oint_C f\,dz=-2\pi\mu\sum_C {\rm res}(f)\,.
\end{eqnarray}
The last line follows because $u$ in Eq.~(\ref{uformula}) is single-valued.

\section{Pressure in Doubly Periodic Flows}
\label{Pressure}
Consider the integer lattice generated by two complex numbers, $\omega_1$ and $\omega_3$, with
$\Im(\omega_3/\omega_1)>0$, consisting of the points
\begin{equation}
W_{mn}=2m\omega_1+2n\omega_3
\end{equation}
for all integers $m$ and $n$.  Suppose identical obstacles are located at these places,
forming a doubly periodic array.  A Stokes flow through this array, represented as in
Eq.~(\ref{uformula}), is characterized by a function $f$ with very restrictive properties.
It is single-valued, it is
holomorphic outside the obstacles, it has poles inside the obstacles, its imaginary part is
doubly periodic, and its real part is a doubly periodic function plus a linear function,
where the linear function is essentially the average pressure $<P>$, increasing linearly along its (constant)
gradient.  If we further ask for the simplest function
of this type, having only simple poles at the $W_{mn}$, then there is essentially only
one possibility, the Weierstrass zeta function $\zeta(z)$, with a linear correction.  This follows
from the theory of elliptic functions \cite{Lawden}.
(Note that $\zeta$ depends also on the lattice constants $\omega_1$
and $\omega_3$, but we regard these as fixed parameters and do not indicate this dependence.)

More precisely we make use of the quasi-periodicity of $\zeta(z)$,
\begin{equation}
\label{quasiperiodic}
\zeta(z+2\omega_\alpha)=\zeta(z)+2\eta_\alpha
\end{equation}
for $\alpha=1,3$, where $\eta_\alpha=\zeta(\omega_\alpha)$ are constants satisfying
\begin{equation}
\label{eta1eta3}
\eta_1\omega_3-\eta_3\omega_1=i\pi/2
\end{equation}
Then we can take, for the function $f$, either of
\begin{equation}
f_\alpha=\frac{-i}{|\omega_\alpha|}[\omega_\alpha\zeta(z)-\eta_\alpha z]\,,
\end{equation}
where $\alpha=1,3$.
Using Eqs.~(\ref{quasiperiodic}) and (\ref{eta1eta3}), we verify that
\begin{eqnarray}
f_1(z+2\omega_1)&=&f_1(z)\\
f_1(z+2\omega_3)&=&f_1(z)-\pi/|\omega_1|\\
f_3(z+2\omega_1)&=&f_3(z)+\pi/|\omega_3|\\
f_3(z+2\omega_3)&=&f_3(z)
\end{eqnarray}
Thus in both functions only the real part, which we interpret as $P/\mu$, shows
a linear growth.  The average pressure gradient corresponding to $f_\alpha$ is
perpendicular to $\omega_\alpha$.  From its projection on the other lattice
vector we determine that it is
\begin{equation}
<\vec{\nabla} P>\quad\leftrightarrow\quad 2<\partial P/\partial\overline{z}>
=\frac{-i\omega_\alpha\mu\pi}{2|\omega_\alpha|\Im (\overline{\omega_1}\omega_3)}
\end{equation}
By taking linear
combinations of $f_1$ and $f_3$ we can find a suitable $f$ with the pressure gradient in
any direction with respect to the lattice.  The combination corresponding to a pressure
gradient of the same magnitude as $f_1$ and $f_3$ but in the direction $e^{i\theta}$ is
\begin{equation}
f=e^{i\theta}\zeta-Cz
\end{equation}
with
\begin{equation}
C=i\left(\frac{\eta_1\Re(e^{-i\theta}\omega_3)-\eta_3\Re(e^{-i\theta}\omega_1)}{\Im(\overline{\omega}_1\omega_3)}\right)
\end{equation}

The force on the fluid in a period parallelogram $D$, with sides $2\omega_1$ and $2\omega_3$, due to the
average pressure in the flow corresponding to $f_\alpha$, is
\begin{eqnarray}
F_P&=&\oint_{\partial D}<P>idz=i\int\int_D <\partial P/\partial\overline{z}>d\overline{z}dz\\
&=&-2<\partial P/\partial\overline{z}>\int\int_D dx\,dy=\frac{2i\omega_\alpha\mu\pi}{|\omega_\alpha|}
\end{eqnarray}

The force on the fluid due to each obstacle in this same flow,
since $\zeta(z)$ is normalized to have residue 1
at each simple pole, is $-2 i\omega_\alpha\mu\pi/|\omega_\alpha|$, by Eq.~(\ref{FonObstacle}).
There is on average
one obstacle in each period parallelogram.  Thus the force on the fluid due to the average pressure
(applied somehow from outside)
is balanced by the force due to the obstacles, and the net force on the fluid is zero, as
is always the case in Stokes flows.

A more general function $f$ in the representation of Eq.~(\ref{uformula}) for the Stokes flow $u$
would be a superposition of translates of $f_1$ and $f_3$, always
keeping the singularities inside the obstacles, and not in the physical region of flow.
Equivalently, one could take a multipole expansion of such functions.  This would be a series in derivatives
of $f_1$ and $f_3$.  These functions are doubly periodic with higher order poles at the
lattice points, that is, they are the Weierstrass $\cal{P}$ function and its derivatives.
Thus $f$ has the form
\begin{equation}
f(z)=e^{i\theta}\zeta-Cz+\sum_n^{N_c} c_n {\cal{P}}^{(n)}
\end{equation}
If the boundary contours have reflection symmetry in the origin, then it is enough to
take only the terms in the series with $n$ odd.

\section{Velocity Field in Doubly Periodic Flows}
From Eq.~(\ref{uformula}) we know that the flow $u$ corresponding to $f$ above is
\begin{equation}
\label{uperiodic}
u=\frac{1}{4}\left(ze^{-i\theta}\overline{\zeta}
-\overline{C}|z|^2
- 2e^{i\theta}\ln |\sigma|-C z^2/2
+z\sum_n^{N_c}\overline{c}_n \overline{{\cal{P}}^{(n)}}
-\sum_n^{N_c}{c_n\cal{P}}^{(n-1)}+\overline{g}\right)
\end{equation}
where $g$ is a second analytic function, still to be determined.  We have already used the
freedom to add terms of the form $\overline{g}$ in adding a term $e^{i\theta}\ln \overline{\sigma}$,
where $\sigma$ is the sigma function, another of the special functions of elliptic function
theory \cite{Lawden}.  Such logarithmic terms compensate the multivaluedness of logarithmic terms
in the integral of $f$.

Now among all the functions $g$ that we could choose, we want the one that makes
$u$ doubly periodic, and that makes $u=0$ (say) on the boundary contours of the obstacles.
Since the logarithmic part of $g$ has been explicitly written as a separate term, $g$ as defined above
is single-valued, and hence has
a Laurent series in a neighborhood of the origin.  We must anticipate, though, that $g$ has singularities
at all the lattice points $W_{mn}$, and hence that this series would converge only out to the nearest
such lattice point.  This difficulty can be removed because
double periodicity of $u$ determines the principal part of $g$ at all lattice
points.  We see, for example, that $u$ blows up logarithmically at $z=0$ but, because
of the poles in $\zeta$, it blows up like $1/z$ at every other lattice point.  This is
inconsistent with double periodicity, and hence $g$ must cancel the $1/z$ behavior
at all $W_{mn}$ except $0$.  Also, in the Laurent series for $g$ about $z=0$, there will
be negative powers of $z$.  Translates of these terms contribute to the principal part
of $g$ at other $W_{mn}$.  Let $S$ be a set of indices $(m,n)$ sufficiently large to label
all $W_{mn}$ in a disk large enough to contain a period parallelogram.  Then we take $g$
in the form
\begin{equation}
\label{gperiodic}
g=-\sum_{(m,n)\in S}\frac{\overline{W}_{mn}}{z-W_{mn}}+\sum_{j=0}^{N_b} b_j\sum_{(m,n)\in S}(z-W_{mn})^{-j}
+\sum_{j=1}^{N_a}a_j z^j
\end{equation}
Because the poles of $g$ have been included as explicit terms, the power series
with coefficients $a_j$ now converges
out to the nearest $W_{mn}$ not indexed in $S$, and we choose this index set large enough
to get good numerical behavior in our final step.

Lastly we specify the boundary conditions that determine the coefficients
$a_n$, $b_n$, and $c_n$ in Eqs.~(\ref{uperiodic}) and (\ref{gperiodic}).  We choose
a region $D$ with a boundary $\partial D$ that is entirely within the fluid region, and such that every
point $z$ in $\partial D$ has a corresponding point $z'$ in $\partial D$
displaced by one of the lattice constants $\pm 2\omega_\alpha$.  Typically $D$ would be the period
parallelogram centered on the origin with vertices $(\pm \omega_1,\pm \omega_3)$.
We ask that the complex function $u$ be periodic, taking the same value at points
$z$ and $z'$ in $\partial D$
related by a lattice constant, and also that the derivative $\partial u/\partial \overline{z}$
be periodic in the same sense.  We also require that $u$ take
prescribed values
on the boundary contour $C$ of the obstacle.  Choosing a large number of pairs of points $(z,z')$
on $\partial D$ and also points $z$ on $C$ for imposing these conditions, we obtain
an overdetermined inhomogeneous system of linear equations that we then solve in the sense of
least squares.  Because $c_n$ occurs together with its complex conjugate $\overline{c_n}$, the system
must be regarded as real linear, not complex linear.
With suitable choices for the truncation parameters $N_a$, $N_b$, and $N_c$, this
boundary value problem has a solution accurate to many decimal places.  We confirm
the solutions in \cite{Acrivos1}, for flows through square and triangular lattices
of obstacles, to 5 decimal
places, typically.

It is superficially surprising that one can stipulate conditions on both $u$
and its derivative in this elliptic problem, but it is clear physically
that such periodic solutions must exist.  It is also clear mathematically, if one considers the
equivalent Stokes flow on a torus.  This argument only applies, though, if the applied
stress that drives the flow has all the properties that we have required of the
function $f$ in Section~\ref{Pressure}.  For other functions $f$ the boundary
value problem would have no solution.  Knowing the form of $f$ in advance, as we do, the actual
boundary values of $P$ and $u$ emerge as part of the solution.  Other approaches
to this problem are typically restricted to situations in which one knows
the boundary values by symmetry, where
$P$ is constant on part of $\partial D$, for example, or where $u$ is normal or tangential on
$\partial D$, special cases that do not hold for general lattices and flows.

Let us define the \emph{mean} of a flow $u$ to be the constant flow $V$ such that
the flux of $V$ and the flux of $u$ into a period parallelogram are the same.
By incompressibility of the flow, this is also the flux of $V$ (and $u$)
out of a period parallelogram.  If we take the period parallelogram to be the
one with vertices $(\pm \omega_1,\pm \omega_3)$, then we have the formula
\begin{equation}
\label{meanV}
V=\frac{1}{2\Im(\overline{\omega_1}\omega_3)}
   \left[\Im\left(\int_{\omega_1-\omega_3}^{\omega_1+\omega_3} \overline{u}\, dz\right)\omega_1
        +\Im\left(\int_{\omega_1+\omega_3}^{-\omega_1+\omega_3} \overline{u}\, dz\right)\omega_3\right]
\end{equation}
The integrals depend only on the endpoints, by incompressibility of the flow, and the
formula is invariant under a common translation of the endpoints.  Thus the choice of period
parallelogram is in fact arbitrary.

\begin{figure}[ht]
\includegraphics{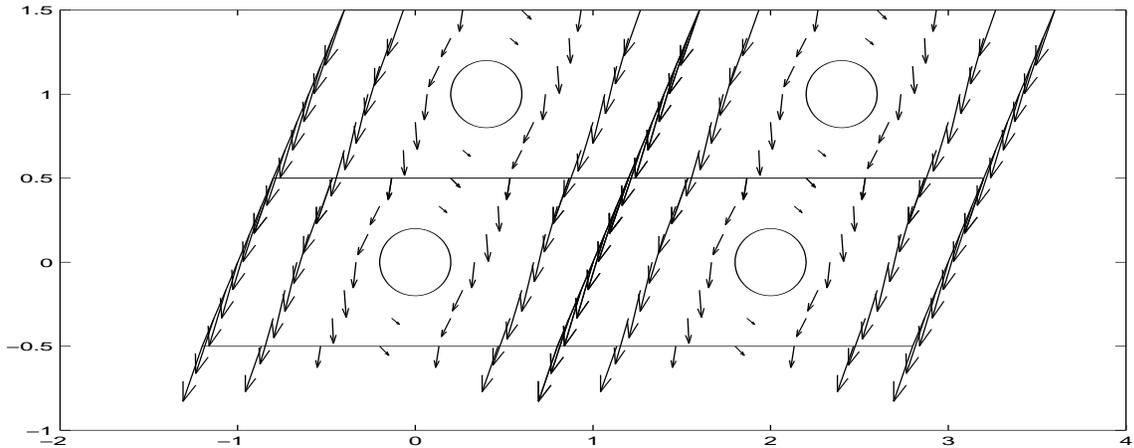}
\caption{A Stokes flow through a lattice of circular obstacles is shown in
four period parallelograms.  The lattice vectors are $\omega_1=1$, $\omega_3=0.2+0.5i$,
and the obstacles have radius $0.2$.
The pressure gradient is vertical, but the mean flow is, to good approximation, along
$\omega_3$, the shorter of the lattice vectors.}
\label{Flow2}
\end{figure}

Using this formulation we have computed the tensorial relation between the force $F$
in Eq.~(\ref{FonObstacle}) and the mean velocity $V$ in Eq.~(\ref{meanV}) in many
examples.
For regular square and triangular lattices, $F$ and $V$ are simply proportional,
but more generally
\begin{equation}
V=GF
\end{equation}
where the conductance tensor $G$ is a real symmetric matrix (and we are now
representing $V$ and $F$ as real 2-vectors, not complex scalars).
A typical computed flow
is shown in Figure~\ref{Flow2}.
Here $\omega_1=1$, $\omega_3=0.2+0.5i$, and the circular obstacles have radius $0.2$.
The pressure gradient $F$ is in the $y$ direction, but the mean flow $V$ is approximately along $\omega_3$, the
shorter of the lattice vectors.  The reason for this behavior is simple, and in accord with
common sense.  The direction of $\omega_3$ is an ``easy direction" for the flow, an
eigenvector of $G$, and the corresponding conductance eigenvalue for the rather wide
``channels" along this direction is large.  The obstacles, being relatively closely spaced along
the direction of $\omega_3$, create ``walls"
along the sides of the channels.  The conductance corresponding to the other,
perpendicular eigenvector, normal to the ``walls", is much smaller than the conductance
along the channels.  The ratio of the two eigenvalues, a measure of the anisotropy of
the conductance tensor, is about $2.545$ in this example,
noticeably more anisotropic than the lattice itself.
It is clear that the conductance anisotropy could be indefinitely large for this lattice
if the round obstacles were larger and almost
touched along the $\omega_3$ direction.

The theory described here provides a starting point for problems involving
flow near ciliated surfaces.  Within a ciliated layer, the cilia, hairlike
projections from the surface, would constitute the regular array of
obstacles.  In the simplest case they would be rigid and perpendicular to
the surface. Consider, for example, a straight pipe of circular cross
section, with such a ciliated inner surface, still leaving a free
cylindrical channel down the center. Suppose the length of the cilia is
much greater than the mean distance between them, and much less than the
radius of the pipe. The Poiseuille flow problem asks for the flow in
response to a pressure difference between one end of the pipe and the
other.  The usual boundary condition says that the flow should be zero on
the walls of the pipe, and this condition still determines the flow in a
thin region near the wall, comparable in thickness to the mean spacing of
the cilia, but within most of the ciliated layer, the mean flow will be
given in the manner described here. Via a similarly thin transition layer,
the Poiseuille flow down the unobstructed central channel matches this
mean flow at its edge. Systematic anisotropy in the placement of the
cilia, including, one must anticipate, a merely statistical anisotropy,
would impart rotation to the fluid channel, as it amounts to a ``rifling"
of the inner surface of the pipe.  Considerations of this kind are
relevant in biological settings where such surfaces are very common. The
cilia as we have just described them are merely passive obstacles, but
often they are active agents driving flows. The methods of this paper
could also be a starting point for treating such phenomena, as we hope to
show in the future.
\section*{Acknowledgements}
We thank the Howard Hughes Medical Institute grant \#52005134, the Mount
Holyoke College Summer Research program, and the Hutchcroft Fund of the
Mount Holyoke College Mathematics Department for support.

\end{document}